\documentclass[12pts,twocoloumn,epsf]{revtex4}
\usepackage{graphicx}
\usepackage{hyperref}
\begin{document}

\title{Unconventional transport properties of an itinerant ferromagnet: EuTi$_{1-x}$Nb$_{x}$O$_3$ ($x$=0.10$-$0.20)}

\author{S. Roy, N. Khan, and P. Mandal}

\affiliation{Saha Institute of Nuclear Physics, HBNI, 1/AF Bidhannagar, Calcutta 700 064, India}
\date{\today}
\begin{abstract}
We report the temperature and magnetic field dependence of resistivity ($\rho$) for single-crystalline EuTi$_{1-x}$Nb$_{x}$O$_3$ ($x$=0.10$-$0.20), an itinerant ferromagnetic system with very low Curie temperature ($T_C$). The detailed analysis reveals that the charge conduction in EuTi$_{1-x}$Nb$_{x}$O$_3$ is extremely sensitive to Nb concentration and dominated by several scattering mechanisms. Well below the $T_C$, where the spontaneous magnetization follows the Bloch's $T^{3/2}$ law, $\rho$ exhibits $T^2$ dependence with a large coefficient $\sim$10$^{-8}$ $\Omega$ cm K$^{-2}$ due to the electron-magnon scattering. Remarkably, all the studied samples exhibit a unique resistivity minimum at $T$$=$$T_{\rm min}$ below which $\rho$ shows logarithmic increment with $T$ (for $T_{\rm C}$$<$$T$$<$$T_{\rm min}$) due to the Kondo scattering of Nb 4$d^1$ itinerant electrons by the localized 4$f$ moments of Eu$^{2+}$ ions which suppresses strongly with applied magnetic field. In the paramagnetic state, $T^{2}$ and $T^{3/2}$  dependence of the resistivity have been observed, suggesting an unusual crossover from a Fermi-liquid  to a non-Fermi-liquid behavior with increasing $T$. The observed temperature and magnetic field dependence of resistivity has been analysed using different  theoretical models.\\
\end{abstract}
\maketitle
\pagebreak
The interplay between magnetism and charge conduction in strongly correlated systems leads to several fascinating physical phenomena [1-3]  such as the unusual temperature dependence of resistivity due to non-Fermi-liquid behavior, Kondo effect, quantum phase transition, complex magnetic excitations, anomalous Hall effect, large negative magnetoresistance etc. [1-16]. Hence any new correlated itinerant ferromagnet immediately calls for exploring its magnetic and transport properties. Based on magnetic and transport properties, metallic ferromagnets  can be divided into two broad categories: i) good metallic ferromagnets such as iron, nickel, and cobalt, which can be described by the Landau Fermi liquid theory [17] and   ii) `bad metallic' ferromagnets such as heavy fermion compounds, transition metal oxides, perovskite manganites, cobaltite, nickelets, etc. [18-29]. For these so-called bad metals often the Fermi liquid description breaks down due to strong electronic correlation. The celebrated itinerant-electron ferromagnet SrRuO$_3$ [1, 30] is an widely studied compound that neither can be classified as a good metallic ferromagnet nor can be compared with 3$d$ transition metal oxides like manganites. Also, the transport and magnetic properties of a very weak itinerant ferromagnet such as ZrZn$_2$ [31] are quite different from those of conventional ferromagnetic (FM) metals.\\

The perovskite titanate family $R$TiO$_{3}$ ($R$=rare-earth ion) with 3$d^1$  electron configuration is one of the fascinating transition-metal oxide systems [32-35]. In last two decades, there is a considerable amount of theoretical and experimental works for understanding the nature of orbital ground state in this family [32-35]. Among the perovskite titanates, EuTiO$_{3}$ is unique because unlike other rare-earth ions, Eu is divalent (4$f^7$) and hence Ti is tetravalent (3$d^0$) [36-45]. Also, EuTiO$_{3}$ is an antiferromagnetic (AFM) ($T_{\rm N} = 5.5$ K) [36-39] band insulator and quantum paraelectric with simple cubic perovskite structure while other titanates are Mott insulator [32-35]. The introduction of electron into the Ti 3$d$ orbital via the substitution of Gd$^{\rm 3+}$ or La$^{\rm 3+}$ ion at the Eu$^{2+}$ sites transforms EuTiO$_{3}$ into a FM  metal [37,39]. The electrical transport and magnetic properties have been reported for thin films and single crystals of Eu$_{1-x}$La$_x$TiO$_3$ (ELTO) with  $x$ up to 0.10 only [37,39]. Except at low temperature, the temperature dependence of resistivity ($\rho$) for ELTO with $x$=0.10 appears to be similar to 10$\%$ La-doped SrTiO$_3$. For Sr$_{0.9}$La$_{0.1}$TiO$_3$ (SLTO), $\rho$ decreases smoothly down to very low temperature without showing any anomaly while a sharp drop in $\rho$ is observed just below the FM transition temperature ($T_C$) in ELTO. Most importantly, ELTO shows a weak upturn in $\rho$ below $\sim$30 K [37,39]. Another very promising candidate is obtained when electrons are introduced in EuTiO$_3$ by Nb$^{4+}$ (4$d^{1}$) ion substitution at the Ti$^{\rm 4+}$ (3$d^0$) site  without breaking the magnetic chains of Eu$^{\rm 2+}$ moments.  A notable increase in electrical conductivity is observed in EuTi$_{1-x}$Nb$_x$O$_{3}$ (ETNO) with the increase in Nb concentration ($x$) [42-46]. In contrast to ELTO, the substitution is done at $B$-site for ETNO and it shows metallic and FM behavior over a much wider range of Nb doping (0.05$<$$x$$\leq$1) [45,46]. Insulator to metal transition in $AB$O$_3$-type perovskite materials, through $B$-site substitution is extremely rare. Normally, $B$-site substitution creates disordering which enhances carrier localization effect.  In spite of some minor differences, the nature of magnetism and transport properties in both metallic ELTO and ETNO are similar. Remarkably, ETNO also exhibits an upturn in $\rho$ below 30 K [45]. Though $\rho$ decreases rapidly with decrease in $T$ for both ELTO and ETNO, it is quite unexpected that the mechanism of charge conduction in these two systems will be same as in SLTO. Due to the presence of large localized moments, there may be strong interaction between the 4$d^1$ itinerant charge carrier and the spin ($S$=7/2) of Eu$^{2+}$.  \\

In order to explain  the unusual temperature dependence of resistivity and shed some light on charge conduction mechanism in EuTi$_{1-x}$Nb$_x$O$_{3}$, we report a comprehensive study of transport properties on high quality single-crystalline samples with  $x$=0.10, 0.15 and 0.20. To the best of our knowledge, the detailed analysis of temperature and magnetic field dependence of resistivity  for  EuTi$_{1-x}$Nb$_x$O$_{3}$ or Eu$_{1-x}$La$_x$TiO$_3$ has not been done so far. The observed results are  compared and contrasted with different classes of FM metals mentioned above. We have also compared the present results with the reported behavior of $\rho$($T$) curve for Eu$_{1-x}$La$_x$TiO$_3$ to know whether the charge conduction mechanism is same for both the systems.  Indeed, our detailed analysis of temperature dependence of resistivity unveils the presence of several unusual scattering mechanisms. With increasing $T$, charge scattering mechanism crosses over from electron-magnon to spin-disordering to Kondo to electron-electron to an unusual $T^{3/2}$ dependence of resistivity due to non-Fermi liquid behavior.\\

{\textbf{Results}}\\
{\textbf{Magnetization.}} Nb doping destabilises the antiferromagnetic  state in EuTiO$_{3}$ and the system becomes FM above a critical doping $x_c$$\sim$0.05. Several theoretical studies also reveal an intricate balance between AFM and FM interactions in EuTiO$_{3}$ [36,38]. $T_C$ is observed to increase sharply with $x$, becomes maximum at around 0.15 and then decreases slowly with further increase of $x$. $T_C$ determined from the low-field magnetization data are 8, 9.5 and 6 K for $x$=0.10, 0.15 and 0.20, respectively [see Supplementary Figure 2]. The value of saturation magnetization at 2 K and 9 T is very close to the expected moment 7 $\mu_B$/Eu. The observed behavior of magnetization is qualitatively similar to ELTO, though the maximum $T_C$ is 1.5 K higher in ETNO [39]. For all the studied ETNO compounds ($x$=0.10$-$0.20), the temperature dependence of specific heat exhibits a distinct $\lambda$-like anomaly at paramagnetic (PM) to FM transition [see Supplementary Figure 3].  We have not observed any anomaly other than that at $T_C$ either in magnetization or in heat capacity data. This suggests that unlike polycrystalline samples [45], the studied crystals are chemically homogeneous and AFM and FM phases do not coexist in the studied composition range. In the PM state, susceptibility obeys the Curie-Weiss law with effective moment, $P_{\rm eff}$=7.9 $\mu_B$/Eu [43].\\

For understanding the correlation between transport and magnetism, the magnetic excitation spectrum has been investigated below $T_C$. The  magnetic excitation is studied by estimating the spontaneous magnetization at different temperatures from the field and temperature dependence of magnetization measured up to 9 T magnetic field. Figure 1(a) shows the temperature dependence of the spontaneous magnetization for the $x$=0.15 sample as a representative.

\begin{figure}
\includegraphics[width=0.5\textwidth]{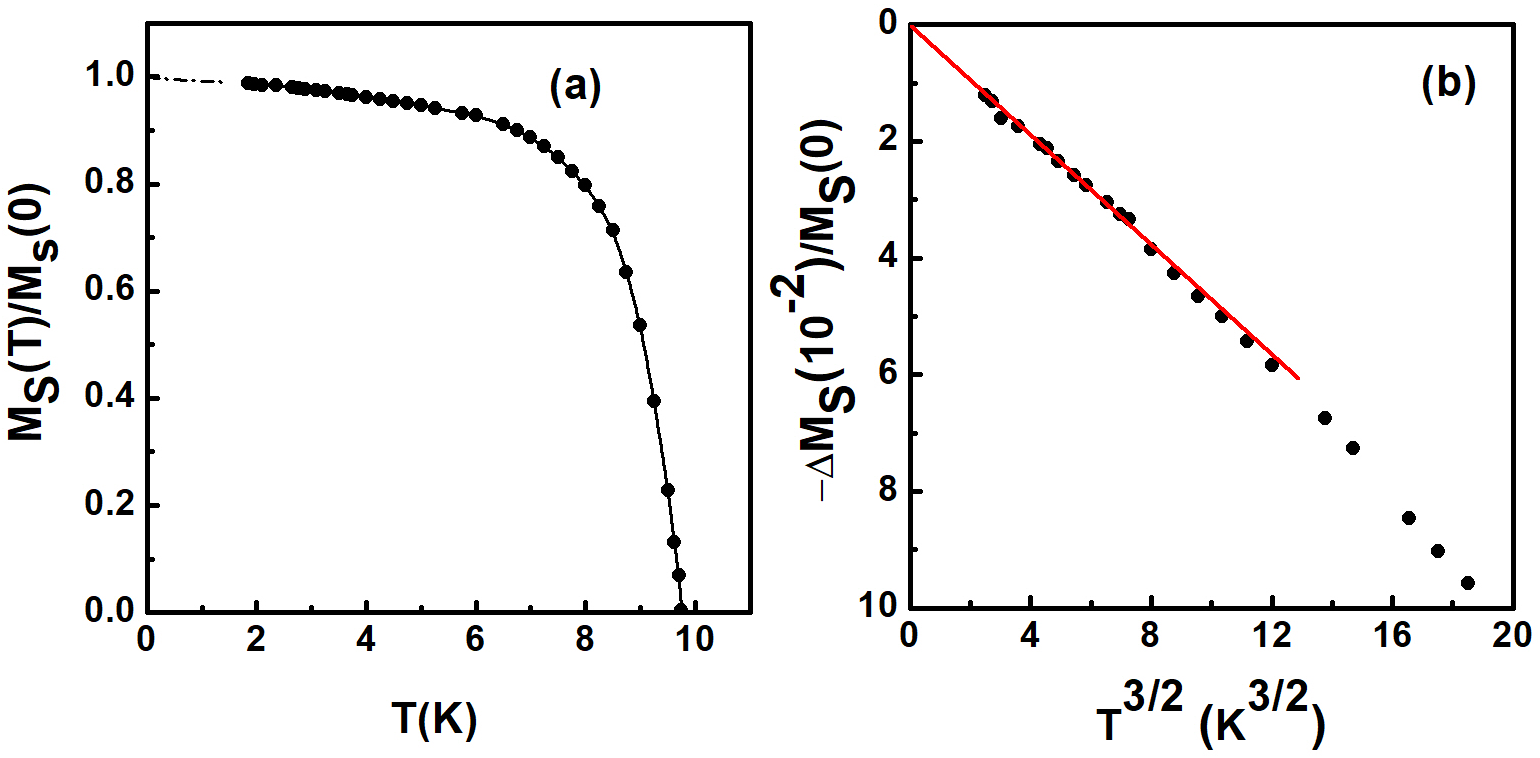}
\renewcommand{\figurename}{\textbf{Figure}}
\caption{\textbf{Magnetization for the EuTi$_{0.85}$Nb$_{0.15}$O$_{3}$  single crystal.} (a) Temperature dependence of the spontaneous magnetization $M_S$($T$) normalized by that at $T$$=$0 K. The solid and dashed lines are guide to the eye. (b) $T^{3/2}$ dependence of the spontaneous magnetization below $T_C$ where the solid red line is the linear fit due to the Bloch's $T^{3/2}$ law.}
\end{figure}

 Well below the transition temperature, where the critical fluctuation is absent, the temperature variation of the spontaneous magnetization can be explained by the Bloch's $T^{3/2}$ law [47],
\begin{equation}
 [M_{\rm S}(0) - M_{\rm S}(T)]/ M_{\rm S}(0) = DT^{3/2},
\end{equation}
where $M_{\rm S}(0)$ and $M_{\rm S}(T)$ are the spontaneous magnetization at $T$=0 and at a finite $T$, respectively and $D$ is the spin-wave parameter. In case of simple pseudo-cubic lattice, the coefficient $D$ is determined by the following relation
\begin{equation}
 D = (0.0587/ S)(k_{B}/2JS)^{1.5},
 \end{equation}
where $S$ is the total spin of Eu$^{\rm 2+}$, $k_{\rm B}$ is the Boltzmann constant and $J$ is the exchange coupling between two neighboring Eu$^{\rm 2+}$ moments.  Figure 1(b) shows that the spontaneous magnetization at low temperature obeys the Bloch's $T^{3/2}$ law. From the slope of the linear fit to the data, we obtain the value of $D$ as 5$\times 10^{-3}$ K$^{\rm -3/2}$. Using the value of $S$=7/2 for Eu$^{\rm 2+}$ spin, from Eq.(2) the strength of exchange interaction is estimated to be 0.3 $k_{\rm B}$. In this context, we would like to mention that the above value of $J$ estimated from the Bloch's $T^{3/2}$ law is comparable to that predicted by the molecular field theory [47], $J$$=$$3k_B T_{C}/2ZS(S+1)$=0.15 $k_{\rm B}$, where $Z$ is the number of nearest neighbor spins. For a simple cubic perovskite structure, $Z$$=$6. \\

{\textbf{Resistivity.}} Figure 2(a) depicts the temperature dependence of $\rho$ at zero-field for the EuTi$_{1-x}$Nb$_x$O$_{3}$  single crystals. Over the entire temperature range, $\rho$ exhibits metallic behavior (d$\rho/d$$T>$0), except in a narrow region just above $T_C$, where a weak increase in $\rho$ has been observed with decreasing $T$ for all the three samples. $\rho$  drops sharply just below $T_{\rm C}$ due to the suppression of spin-disorder scattering. In the FM state, $\rho$  decreases at a much faster rate with decreasing $T$ as compared to that in PM state. Though the qualitative nature of $\rho$($T$) curve is similar to earlier report on polycrystalline samples, for a given Nb content, the absolute value of $\rho$ is smaller and the residual resistivity ratio $\rho$(300 K)/$\rho$(2 K) is much larger in single crystals. For example, the values $\rho$(2 K) and $\rho$(300 K)/$\rho$(2 K) are 150 $\mu$$\Omega$ cm and 1.5 respectively for polycrystalline sample [45] while the corresponding values are 12 $\mu$$\Omega$ cm and 8 for the single crystal, with $x$$=$0.20. At low temperature, the large value of $\rho$ in polycrystalline samples is due to the strong grain boundary scattering.\\

\begin{figure*}
\includegraphics[width=0.9\textwidth]{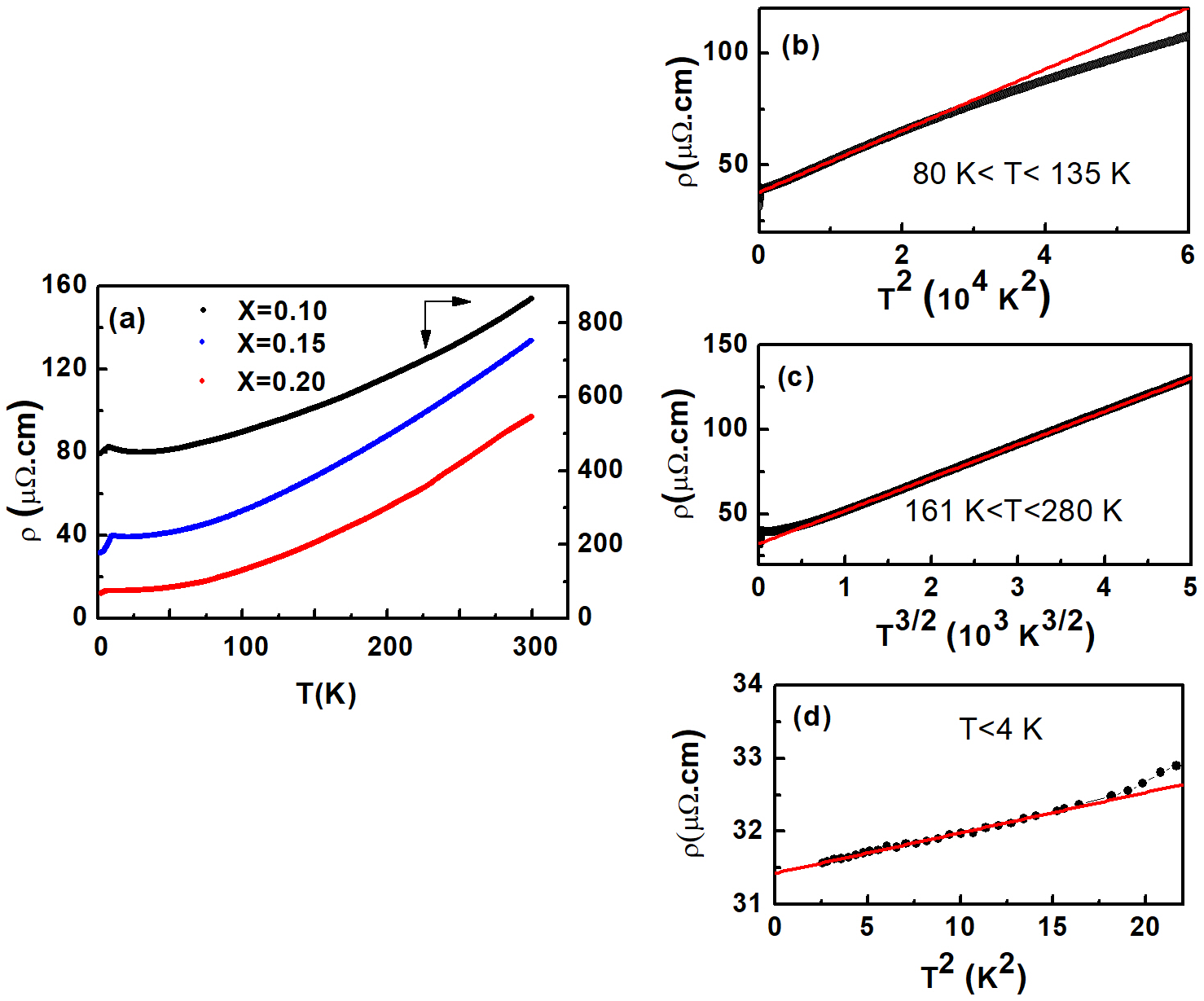}
\renewcommand{\figurename}{\textbf{Figure}}
\caption{\textbf{Resistivity.} (a) Temperature dependence of zero-field resistivity for the EuTi$_{1-x}$Nb$_x$O$_{3}$ ($x$$=$0.10, 0.15, and 0.20) single crystals. The right handed axis and bottom axis scale for the $x$$=$0.10 compound (the black dots) whereas the left handed axis and bottom axis scale for the other two compounds $x$$=$0.15 (the blue dots), $x$$=$0.20 (the red dots). For all the three curves, the PM-FM transition is reflected in resistivity. (b) $T^{2}$ dependence of resistivity in the paramagnetic region 80 K $<$ $T$ $<$ 135 K for the $x$$=$0.15 crystal.  (c) $T^{3/2}$ variation of resistivity for the $x$$=$0.15 crystal in the temperature range of 161 K $<$ $T$ $<$ 280 K. (d) $T^{2}$ fitting to the resistivity at very low temperature in the FM region below 4 K for the $x$$=$0.15 crystal.}
\end{figure*}

From figure 2(a), it is clear that the metallic behavior of $\rho$ in the PM state cannot be fitted with a single power-law expression over the entire temperature range, because  before $\rho$  starts to increase, a strong upward curvature develops at low temperature. To explain the intricate nature of temperature dependence of $\rho$, the role of different scattering mechanisms on charge conduction has been analyzed in details. In the PM state, $\rho$ exhibits a quadratic $T$ dependence, $\rho$$=$$\rho_{a0}$ + $aT^2$, in a narrow range above the resistivity minimum followed by a crossover to $\rho$$=$$\rho_{b0}$ + $bT^{3/2}$ dependence over a wider temperature range and up to $T$ as high as 280 K. The observed behavior of $\rho$ is demonstrated in Figures 2(b) and 2(c) for $x$=0.15 as a representative. The deduced values of $a$ and $b$ in two different temperature ranges in the PM state are presented in Table I for all the three samples [see Supplementary Figures 4 \& 5]. The $T^2$ behavior of $\rho$ is an indication of electronic correlation, consistent with the formation of  Fermi-liquid state. The coefficient $a$ is a measure of the quasiparticle-quasiparticle  scattering rate. As expected, $a$ decreases rapidly with the increase in carrier doping. Often, metallic oxides  are termed as 'bad metals' due to their strong electron-electron interaction. The deduced value of $a$ for $x$=0.20 is about an order of magnitude smaller than that for the well known itinerant FM SrRuO$_3$ but two orders of magnitude larger than that for the elemental FMs such as Fe, Co, Ni [28].\\

It is noteworthy  that $\rho$ in several FM perovskites shows $T^2$ dependence. However, the $T^2$ dependence in these systems has been observed well below the $T_{\rm C}$. $\rho$ also exhibits $T^2$ dependence well below $T_{\rm C}$ for ETNO, as shown in Figure 2(d) for $x$=0.15. Contrary to the observed $T^2$ behavior of $\rho$ in the PM state, $T^2$ behavior in the FM state results in unusually large value of the coefficient $a$, estimated to be about two orders of magnitude larger than that in PM state. Such a  huge difference in the values of $a$ in PM and FM states indicates that the $T^2$ dependence of $\rho$ in the FM state is due to the dominant magnetic scattering over the electron-electron scattering. In ferromagnets, $\rho$ shows $T^2$ dependence at low temperature due to the electron-magnon scattering. As both electron-electron and electron-magnon scattering occur at low temperature, it is very difficult to separate their relative contributions in resistivity for FM materials with high Curie temperature.
Even in elemental FM, where the electronic correlation is believed to be very weak, the origin of $T^2$ behavior of $\rho$ at low temperature is not yet settled [48]. Due to the much lower PM to FM transition temperature of the studied system, we have been able to detect both the scattering in two different temperature regions. \\

\begin{table*}
{
\caption{Estimated values of the resistivity coefficient $a$ in units of $\mu$$\Omega$ cm/K$^2$ in the FM ($T<T_{\rm C}$) and PM ($T>T_{\rm C}$) states and the resistivity coefficient $b$ in units of $\mu$$\Omega$ cm/K$^{3/2}$ in the PM state ($T>T_{\rm C}$) for EuTi$_{1-x}$Nb$_x$O$_{3}$ single crystals with $x$=0.10, 0.15, and 0.20. For $x$$=$0.20, the accurate determination of the parameter $a$ below $T_{\rm C}$ is not possible due to its low transition temperature.}\label{I}
\begin{tabular*}{1.0\textwidth}{@{\extracolsep{\fill}}c| c c |c c |c c }
\hline
\hline
$x$ & $a$ ($T<T_{\rm C}$)  & Range of $T$ & $a$ ($T>T_{\rm C}$) & Range of $T$ & $b$ ($T>T_{\rm C}$) & Range of $T$ \\
\hline
0.10 & 26.9(1)$\times$10$^{-2}$ & T$<$2.9 K & 6.0(1)$\times$10$^{-3}$ & 73$<T<$120 K & 8.4(1)$\times$10$^{-2}$ & 175$<T<$260 K \\
\hline
0.15 & 5.5(5)$\times$10$^{-2}$ & $T<$4 K & 1.4(1)$\times$10$^{-3}$ & 80$<T<$135 K & 2.0(1)$\times$10$^{-2}$ & 161$<T<$280 K \\
\hline
0.20 &  &  & 1.1(1)$\times$10$^{-3}$ & 85$<T<$130 K & 1.7(1)$\times$10$^{-2}$ & 150$<T<$233 K \\
\hline
\hline
\end{tabular*}}
\end{table*}
At high temperature, resistivity shows marked departure from the usual temperature dependence of a conventional metal. Normally, resistivity of a  metal  shows linear $T$ dependence at high temperatures due to the electron-phonon scattering.  However, unconventional $T^{3/2}$  behavior of $\rho$  has been observed in some magnetic systems. For a three-dimensional AFM system, $T^{3/2}$ dependence of resistivity is expected to occur in the vicinity of the AFM transition due to the spin fluctuations. In systems like Pd-based dilute alloys such PdFe, PdMn (with few $\%$ of Fe or Mn), the observed $T^{3/2}$  dependence of $\rho$ well below the AFM/FM transition temperature is attributed to the incoherent part of the electron-magnon scattering [49]. Thus, $T^{3/2}$  behavior of $\rho$ well above the FM transition for the present ETNO system can not be interpreted in terms of incoherent electron-magnon scattering. The transport properties of undoped rare-earth nickelets $R$NiO$_3$, another strongly correlated system with single valance Ni$^{3+}$, are of great interest in recent time [22,23,50]. In both single crystals and thin films,  $\rho$ exhibits very unusual $T$ dependence. $\rho$ follows $T^{n}$   (with $n$$=$1.33 and 1.6) behavior in single crystals of PrNiO$_3$ under high pressure whereas in ultrathin epitaxial film of NdNiO$_3$, $\rho$  is extremely sensitive to lattice strain and for highly compressive strained-films, $\rho$ shows $T^{5/3}$  dependence [22,23]. However, in La-doped NdNiO$_3$, $\rho$ shows $T^{3/2}$ behavior [50]. $T^{3/2}$ dependence of $\rho$ under high pressure has also been observed in the well known skyrmion lattice MnSi [25,26].  The strong deviation from $T^2$ dependence of $\rho$ in these systems has been attributed to the non-Fermi liquid state. $T^{3/2}$ behavior of $\rho$  has also been reported in single crystal and thin films of itinerant ferromagnet SrRuO$_3$ [28,51]. In SrRuO$_3$, $\rho$ exhibits $T^2$ dependence  well below  $T_C$ ($\sim$160 K) and a crossover  to $T^{3/2}$  dependence occurs with increasing temperature. Similar to ETNO, $T^{3/2}$ behavior is observed over a wide temperature range. This crossover from $T^{2}$ to $T^{3/2}$  dependence  of $\rho$ in SrRuO$_{3}$ has been attributed to the non-Fermi liquid transition [51]. The values of the coefficient $b$ deduced from the $\rho$ vs $T^{3/2}$ plot are comparable to that reported for the SrRuO$_3$ thin film [51].  In SrRuO$_{3}$, the $T^{3/2}$  dependence  in $\rho$ has been observed in the temperature range 50$-$120 K, but the temperature that corresponds to the characteristic magnon energy ($k_{\rm B}T_{\rm m}$) was estimated to be $T_{\rm m}$=70 K. Therefore, for the SrRuO$_3$ compound, the electron-magnon scattering alone can not describe the $T^{3/2}$ behavior and has been attributed to the scattering of itinerant electrons by fluctuation-induced localized electrons. The value of resistivity exponent $n$ is very sensitive to the underlying critical phenomena giving rise to the critical fluctuations associated
with non-Fermi liquid behavior. The widely observed critical exponent 3/2 indicates finite wavevector fluctuations [22].\\

\begin{figure}
\includegraphics[width=0.5\textwidth]{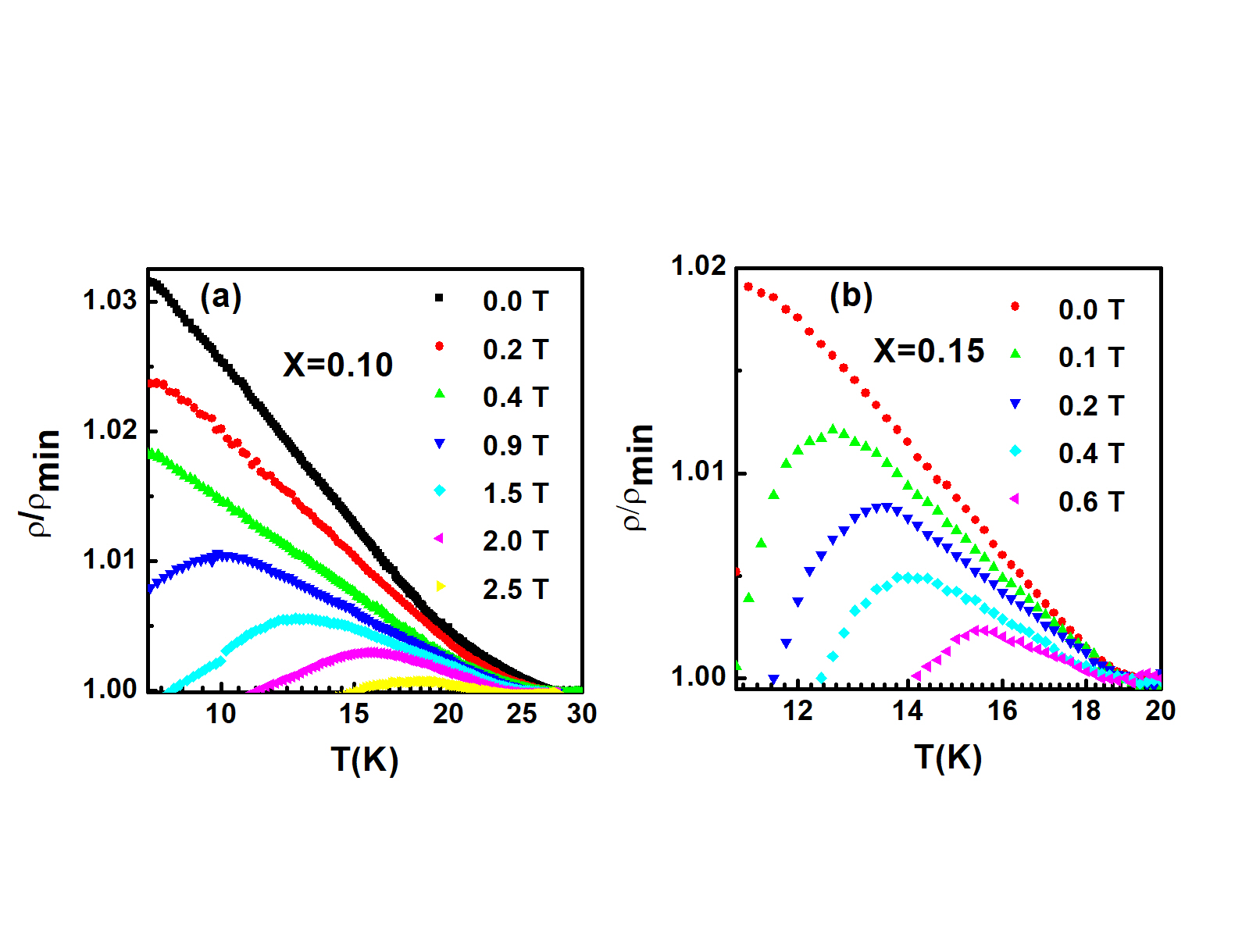}
\renewcommand{\figurename}{\textbf{Figure}}
\caption{\textbf{Kondo-scattering in resistivity.} A broad minimum in resistivity in the paramagnetic region is observed near 30 K and 20 K for the compounds $x$$=$0.10 and $x$$=$0.15 respectively. (a) and (b) show the  log($T$) dependence of the resistivity in the temperature window $T_{\rm C}<T<T_{min}$ at different magnetic fields for the $x$$=$0.10 and $x$$=$0.15 compounds, respectively due to the Kondo scattering. It is clear that resistivity suppresses and log($T$) range shrinks progressively with increasing magnetic field for both $x$$=$0.10 and 0.15.}
\end{figure}

We have already mentioned that $\rho$($T$) curve for all three compositions displays a broad minimum at $T_{\rm min}$ in the PM state [Figure 2(a)] and in the temperature window $T_{\rm C}$$<T$$<T_{min}$, $\rho$($T$) is observed to show log($T$) dependence as shown in Figures 3(a) and 3(b). The effect of magnetic field on the log($T$) dependence of $\rho$ has also been investigated and shown in Figures 3(a) and 3(b) for $x$=0.10 and 0.15, respectively. For $x$$=$0.20, the increase of $\rho$ below $T_{min}$ is very small and as a result, the fit is insensitive. Figures 3(a) and 3(b) show that with the increase in magnetic field, $\rho$ decreases rapidly, the minimum shifts toward lower temperature and the temperature range of log($T$) behavior shrinks progressively. This clearly indicates that the upturns in the resistivity below $T_{min}$ involve the presence of a Kondo-type scattering of the itinerant electrons with the localized 4$f$ spins of the Eu$^{2+}$ ions. It has been suggested that the ferromagnetism in the ELTO and ETNO systems is mediated by the Ruderman-Kittle-Kasuya-Yosida (RKKY) exchange interaction between the itinerant $d$ electrons of Ti or Nb and the localized 4$f$ moments of Eu$^{\rm 2+}$ [37,39,45]. The oscillatory nature of the RKKY interaction which depends on the spatial distance between the localized moments and the Fermi wave vector of the itinerant electrons is believed to be responsible for the decrease of $T_{\rm C}$ for the $x$$=$0.20 compound. The competition of the intersite RKKY exchange interaction and the Kondo effect should result in the formation of either the usual magnetic ordering with large atomic magnetic moments viz., in elemental rare-earth metals or the non-magnetic Kondo state with suppressed magnetic moments [52,53]. The dominant RKKY-type FM exchange below $T_{\rm C}$ stabilizes the usual magnetic ordering with full moment per Eu$^{2+}$ ions in the ETNO system. The low temperatures and high fields favor the FM phase and prevent the formation of Kondo singlets, thereby the temperature range of log($T$) behavior shrinks as the FM phase intrudes into the high-$T$ phase with increasing magnetic field. Therefore, the observed log($T$) dependence of $\rho$ in the present ETNO system mimics the behavior of FM Kondo lattice which is relatively less numerous and exhibits  rather complicated physical pictures [52,53].  In this context, we would also like to mention that few Eu-based compounds show Kondo effect but they order antiferromagnetically [54,55]. \\

{\textbf{Magnetoresistance.}} In order to investigate the effect of the FM ordering of Eu$^{\rm 2+}$ moments on charge scattering, we have measured the temperature dependence of the resistivity for different applied magnetic fields $H$ up to 9 T in the vicinity of and well above $T_{\rm C}$ as shown in Figure 4(a) for the $x$=0.15 compound as a representative. The sharp anomaly in the zero-field resistivity at the FM transition temperature progressively weakens with increasing magnetic field and disappears above a critical field. As a result $\rho$ shows metallic behavior over the entire range of $T$ at high fields. This shows that the conduction of itinerant electrons is strongly coupled with localized Eu$^{\rm 2+}$ spins and can be controlled by aligning the localized spins with magnetic field, and such a process gives rise to large negative magnetoresistance. Figure 4(b) shows the temperature dependence of magnetoresistance (MR), $\Delta$$\rho$$/$$\rho$$=$[($\rho(H)- \rho(0)$)$/$$\rho(0)$] where $\rho(H)$ is the resistivity at a field $H$, for EuTi$_{0.85}$Nb$_{0.15}$O$_3$.  Except in the vicinity of $T_{\rm C}$, $\Delta$$\rho$/$\rho$ is small and decreases rapidly on the both sides of the peak. The maximum value of $\Delta$$\rho$/$\rho$ at 9 T is estimated to be about 21 $\%$.\\

The field dependence of the magnetoresistance $\Delta$$\rho$$/$$\rho$  up to 9 T at different temperatures is shown in Figure 4(c). For clarity only a few representative plots are given. At low temperatures, well below the FM transition, MR is small.  The value of MR increases slowly as the temperature increases, becomes maximum  at around the FM transition and then decreases with further increase of temperature. MR in the FM state exhibits a typical Brillouin-like curvature with magnetic field. These behavior suggest strong correlation between transport properties and magnetic order parameter as observed in the case of manganites and other FM metals.\\
\begin{figure*}
\includegraphics[width=0.9\textwidth]{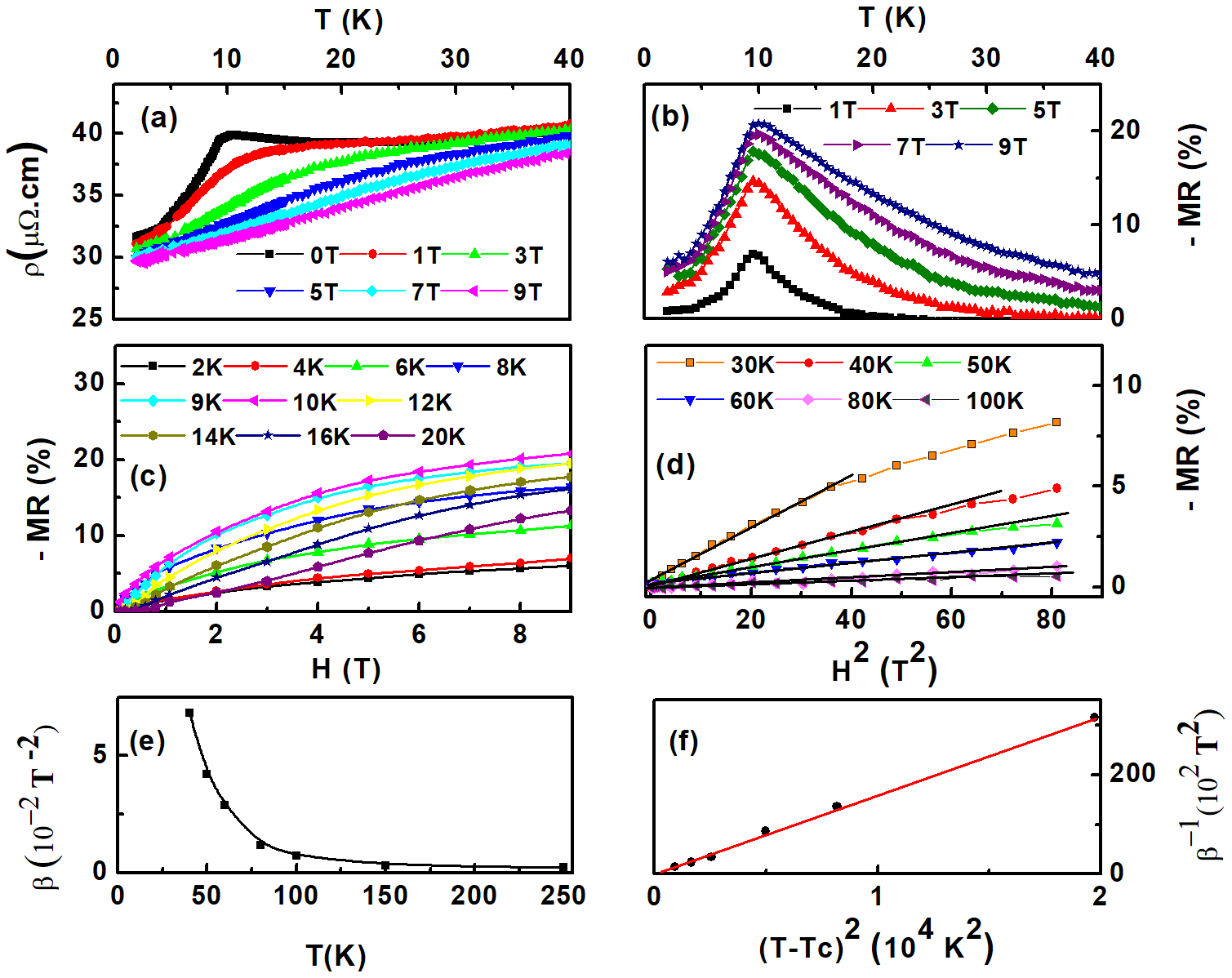}
\renewcommand{\figurename}{\textbf{Figure}}
\caption{\textbf{Magnetoresistance for the EuTi$_{0.85}$Nb$_{0.15}$O$_3$ crystal.} (a) Temperature dependence of the resistivity for different applied magnetic fields. (b) Magnetoresistance (MR) determined from figure (\textbf{a}) is plotted as a function temperature. (c) The field dependence of the magnetoresistance $\Delta$$\rho$/$\rho$ up to 9 T. (d) $H^{\rm 2}$ dependence of the MR well above the $T_{\rm C}$ at low-magnetic fields, where the solid black lines correspond to linear fit. (e) $\beta$ versus $T$ plot. (f) Linear fitting of $\beta ^{-1}$ versus $(T-T_{\rm C})^{2}$ in the paramagnetic region well above $T_{\rm C}$.}
\end{figure*}

The origin of MR in the PM state well above the $T_{\rm C}$ can be understood from the relation of MR with the field induced isothermal magnetization $M$($H$) as observed in several perovskite FMs, i.e.,

\begin{equation}
 MR = [\rho(H) - \rho(0)]/ \rho(0) = - C [M(H)/M_{max}]^{2},
\end{equation}
where $C$ is a constant coefficient and $M_{\rm max}$ is the magnetization at the lowest temperature in the presence of magnetic field $H$ [1]. Therefore, $MR \propto M^{2}$. This relation comes into play as the carriers get scattered by the thermally fluctuating spins.  As the magnetization increases, the fluctuation of spins decreases leading to a decrease in $\rho(H)$, i.e., MR is negative. Temperature dependence of MR is plotted in a wide range below and above $T_{\rm C}$ which shows a peak near $T_{\rm C}$. From Eq.(3) and  $M$$=$$\chi H$ relation, it can be shown that $MR$$=$$-\beta H^{2}$. According to the Curie-Weiss law of paramagnetism, $\chi$ is proportional to $1/(T-T_{\rm C})$ in the PM region well above $T_{\rm C}$. Therefore, $\beta \propto 1/(T - T_{\rm C})^{2}$. Here, the $H^{\rm 2}$ dependence of negative magnetoresistance holds only at low fields but this dependence deviates at higher fields which can be seen in Figure 4(d). This deviation occurs mainly due to the non-linear dependence of magnetization on $H$ near $T_{\rm C}$. $\beta$ versus $T$ and $\beta ^{-1}$ versus $(T-T_{\rm C})^{2}$ have been plotted in Figure 4(e) and Figure 4(f), respectively. The linear fitting of $\beta^{-1}$ indicates that magnetoresistance originates from the local spin fluctuations.\\

{\textbf{Discussion}}\\
When Eu$^{2+}$ is replaced by La$^{3+}$ or R$^{3+}$, electrons are introduced in the three-fold degenerate  3$d$ $t_{2g}$ orbitals of Ti which leads to the mixed valence Ti$^{3+}$ and Ti$^{4+}$ state in the system and the metallic conductivity arises  due to the hopping of electrons from 3$d^1$ to 3$d^0$. This phenomenon is very similar to the evolution of charge conduction with increase of La doping in SrTiO$_3$ and several divalent doped metallic perovskites. In this context, we would like to mention that $\rho$ for Sr$_{1-x}$La$_x$TiO$_{3}$ shows $T^{2}$ dependence due to electronic correlation over a wide range of doping ($x$) and temperature up to 300 K. In $AB$O$_3$ perovskites, $A$ site acts as a charge reserver. When the substitution is made at $A$ site by some heterovalent element, either hole or electron is transferred from $A$ to transition metal ion at $B$ site and the system becomes metallic above a critical doping level.   On the other hand, insulator to metal transition through $B$-site substitution is very rare because substitution at this site creates disordering which favors localization effect. Thus the insulator to metal transition due to the substitution of a very small amount of Nb ($\sim$5\%) at Ti site in ETNO is an unusual phenomenon. The occurrence of metallic conductivity with such a small amount Nb doping means charge conduction is not due to the percolation through Nb but also between Nb and Ti ions. The unpaired 4$d^1$ electron of Nb hops to empty 3$d^0$ orbital of Ti and as a result, 4$d$ state of Nb$^{4+}$ strongly hybridizes with 3$d$ of Ti$^{4+}$. This also suggests that the two energy levels, 3$d$ of Ti and 4$d$ of Nb, are close to each other. At higher doping level $x$ close to 1, Nb is surrounded by other Nb ions, and in such case conductivity is dominated by the hopping of electrons between isovalent Nb ions and makes EuNbO$_3$ metallic. Unlike EuNbO$_3$, the end members of EuTiO$_3$ are Mott insulators when substitution is done at Eu site with other rare earth ion.\\

In summary, we report a comprehensive study of the transport properties in single-crystalline EuTi$_{1-x}$Nb$_x$O$_{3}$ itinerant ferromagnets. The observed $T^{2}$ temperature dependence of the resistivity in the FM state is found to be due to the dominant electron-magnon scattering unlike that observed in the PM state which is due to the electron-electron scattering. Also, a crossover from a $T^{2}$ to $T^{3/2}$ temperature dependence of the resistivity has been observed in the PM state, suggesting a non-Fermi liquid behavior. Furthermore, the scattering of the itinerant electrons by the large localized moments of the Eu$^{2+}$ ions results in a Kondo-like upturn in the resistivity just before the system enters into the FM ordered state. The presence of several scattering mechanisms and the non-Fermi liquid behavior make ETNO a unique ferromagnetic metallic system.\\

{\textbf{Methods}}\\
{\scriptsize Polycrystalline EuTi$_{1-x}$Nb$_x$O$_{3}$ ($x$=0.10$-$0.20) powder samples were prepared by the standard solid-state reaction method. Stoichiometric mixtures of  Eu$_{2}$O$_{3}$ (pre-heated), Nb$_{2}$O$_{5}$ and TiO$_{2}$ were heated at 1000$-$1100$\,^{0}{\rm C}$ for few days in a reduced atmosphere containing 5$\%$ H$_2$ and 95$\%$  argon followed by intermediate grindings. The obtained powder was pressed into two cylindrical rods which are then sintered at 1100$\,^{0}{\rm C}$ in the same environment.  The single crystal was grown from these rods by floating-zone technique using four-mirror image furnace (Crystal Systems) in the reduced atmosphere. The typical growth rate was 5 mm/h. The x-ray diffraction pattern of the powdered single crystal reveals that the compound is of single phase [see Supplementary Figure 1]. Temperature and field dependence of the electrical resistivity was measured by a standard four-probe technique where electrical contacts were made using conducting silver paste. The field dependence of the dc magnetization were performed using both physical property measurement system and SQUID-VSM (Quantum Design).}\\

{\textbf{References}}\\
{\scriptsize
1. Imada, M., Fujimori, A.\& Tokura, Y. Metal-insulator transitions. \emph{Reviews of Modern Physics} \textbf{70}, 1039 (1998).

2. Morosan, E., Natelson, D., Nevidomskyy, A. H. \& Si, Q. Strongly Correlated Materials, \emph{Adv. Mater.} \textbf{24}, 4896 (2012).

3. Klein, L., Dodge, J. S., Ahn, C. H., Reiner, J. W., Mieville, L., Geballe, T. H., Beasley, M. R. \& Kapitulnik, A. Transport and magnetization in the badly metallic itinerant ferromagnet SrRuO$_{3}$. \emph{J. Phys.: Condens. Matter} \textbf{8}, 10111$-$10126 (1996).

4. Doiron-Leyraud, N., Walker, I. R., Taillefer, L., Steiner, M. J., Julian, S. R. \& Lonzarich, G. G. Fermi-liquid breakdown in the paramagnetic phase of a pure metal. \emph{Nature} \textbf{425}, 595$-$598 (2003).

5. Cao, G., Durairaj, V., Chikara, S., DeLong, L. E., Parkin, S. \& Schlottmann, P. Non-Fermi-liquid behavior in nearly ferromagnetic SrIrO$_{3}$ single crystals. \emph{Physical Review B} \textbf{76}, 100402(R) (2007).

6. Stewart, G. R. Non-Fermi-liquid behavior in d- and f-electron metals. \emph{Reviews of Modern Physics} \textbf{73}, 797$-$855 (2001).

7. Pasupathy, A. N., Bialczak, R. C., Martinek, J., Grose, J. E., Donev, L. A. K., McEuen, P. L., Ralph, D. C. The Kondo Effect in the Presence of Ferromagnetism. \emph{Science} \textbf{306}, 86$-$89 (2004).

8. Kharel, P., Skomski, R., Lukashev, P., Sabirianov, R. \& Sellmyer, D. J. Spin correlations and Kondo effect in a strong ferromagnet. \emph{Physical Review B} \textbf{84}, 014431 (2011).

9. Uhlarz, M., Pfleiderer, C. \& Hayden, S. M. Quantum Phase Transitions in the Itinerant Ferromagnet ZrZn$_{2}$. \emph{Physical Review Letters} \textbf{93}, 256404 (2004).

10. Vojta, M. Quantum phase transitions. \emph{Rep. Prog. Phys.} \textbf{66}, 2069$-$2110 (2003).

11. Kirkpatrick, T. R. \& Belitz, D. Universal low-temperature tricritical point in metallic ferromagnets and ferrimagnets, \emph{Physical Review B} \textbf{85}, 134451 (2012).

12. Brando, M., Belitz, D., Grosche, F. M. \& Kirkpatrick, T. R. Metallic quantum ferromagnets, \emph{Reviews of Modern Physics} \textbf{88}, 025006 (2016).

13. Grigera, S. A., Perry, R. S., Schofield, A. J., Chiao, M., Julian, S. R., Lonzarich, G. G., Ikeda, S. I., Maeno, Y., Millis, A. J., Mackenzie, A. P. Magnetic Field-Tuned Quantum Criticality in the Metallic Ruthenate Sr$_3$Ru$_2$O$_7$, \emph{Science} \textbf{294}, 329$-$332 (2001).

14. Kondo, J. Anomalous Hall Effect and Magnetoresistance of Ferromagnetic Metals, \emph{Progress of Theoretical Physics} \textbf{27}, 772$-$792 (1962).

15. Yelland, E. A. \& Hayden, S. M. Magnetic Excitations in an Itinerant Ferromagnet near Quantum Criticality, \emph{Physical Review Letters} \textbf{99}, 196405 (2007).

16. Shitade, A. \& Nagaosa, N. Anomalous Hall Effect in Ferromagnetic Metals: Role of Phonons at Finite Temperature, \emph{Journal of the Physical Society of Japan} \textbf{81}, 083704 (2012).

17. Nozieres, P. \emph{Theory Of Interacting Fermi Systems} (Benjamin, New York) 1964.

18. Smith, R. P., Sutherland, M., Lonzarich, G. G., Saxena, S. S., Kimura, N., Takashima, S., Nohara, M. \& Takagi, H. Marginal breakdown of the Fermi-liquid state on the border of metallic ferromagnetism. \emph{Nature} \textbf{455}, 1220$-$1223 (2008).

19. Tokura, Y. \& Nagaosa, N. Orbital Physics in Transition-Metal Oxides, \emph{Science} \textbf{288}, 462 (2000).

20. Zhang, J., Xu, Y., Cao, S., Cao, G., Zhang,Y. \& Jing, C. Kondo-like transport and its correlation with the spin-glass phase in perovskite manganites, \emph{Physical Review B} \textbf{72}, 054410 (2005).

21. Jaramillo, R., Ha, S. D., Silevitch, D. M. \& Ramanathan, S. Origins of bad-metal conductivity and the insulator-metal transition in the rare-earth nickelates, \emph{Nature Physics} \textbf{10}, 304 (2014).

22. Liu, J., Kargarian, M., Kareev, M., Gray, B., Ryan, P. J., Cruz, A., Tahir, N., Chuang, Yi-De, Guo, J., Rondinelli, J. M., Freeland, J. W., Fiete, G. A. \& Chakhalian, J. Heterointerface engineered electronic and magnetic phases of NdNiO$_3$ thin films. \emph{Nature Communications} \textbf{4}, 2714 (2013).

23. Zhou, J.-S., Goodenough, J. B. \& Dabrowski, B. Pressure-induced non-Fermi liquid behavior of PrNiO$_3$. \emph{Physical Review Letters} \textbf{94}, 226602 (2005).

24. Si, Q. \& Steglich, F. Heavy Fermions and Quantum Phase Transitions, \emph{Science} \textbf{329}, 1161 (2010).

25. Pfleiderer, C., Julian, S. R. \& Lonzarich, G. G. Non-Fermi-liquid nature of the normal state of itinerant-electron ferromagnets. \emph{Nature} \textbf{414}, 427$–$430 (2001).

26. Pfleiderer, C., Reznik, D., Pintschovius, L., L$\ddot{o}$hneysen, H.v., Garst, M. \& Rosch, A. Partial order in the non-Fermi-liquid phase of MnSi. \emph{Nature} \textbf{427}, 227 (2004).

27.  Custers, J., Gegenwart, P., Wilhelm, H., Neumaier, K., Tokiwa, Y., Trovarelli, O., Geibel, C., Steglich, F., Pépin, C. \& Coleman, P. The break-up of heavy electrons at a quantum critical point. \emph{Nature} \textbf{424}, 524$–$527 (2003)

28. Klein, L., Dodge, J. S., Ahn, C. H., Snyder, G. J., Geballe, T. H., Beasley, M. R. \& Kapitulnik, A. Anomalous Spin Scattering Effects in the Badly Metallic Itinerant Ferromagnet SrRuO$_3$. \emph{Physical Review Letters} \textbf{77}, 2774 (1996).

29. S. Sachdev, Quantum Phase Transitions (Cambridge University Press, Cambridge, 1999).

30. Koster, G., Klein, L., Siemons, W., Rijnders, G., Dodge, J. S., Eom, Chang-Beom, Blank, D. H. A., Beasley, M. R. Structure, physical properties, and applications of SrRuO$_3$ thin films, \emph{Reviews of Modern Physics} \textbf{84}, 253 (2012).

31. Yelland, E. A., Yates, S. J. C., Taylor, O., Griffiths, A., Hayden,S. M. \& Carrington, A. Ferromagnetic properties of ZrZn$_{2}$, \emph{Physical Review B} \textbf{72}, 184436 (2005).

32. Zhao,  Z. Y., Khosravani, O., Lee, M., Balicas, L., Sun, X. F., Cheng, J. G., Brooks, J., Zhou, H. D. \&  Choi, E. S. Spin-orbital liquid and quantum critical point in Y$_{1-x}$La$_{x}$TiO$_3$, \emph{Physical Review B} \textbf{91}, 161106(R) (2015).

33. Khaliullin, G. \& Okamoto, S. Quantum Behavior of Orbitals in Ferromagnetic Titanates: Novel Orderings and Excitations, \emph{Physical Review Letters} \textbf{89}, 167201 (2002).

34. Hemberger, J., Krug von Nidda, H.-A., Fritsch, V., Deisenhofer, J., Lobina, S., Rudolf, T., Lunkenheimer, P., Lichtenberg, F., Loidl, A.,
Bruns, D. \& B$\ddot{u}$chner, B. Evidence for Jahn-Teller Distortions at the Antiferromagnetic Transition in LaTiO$_3$. \emph{Physical Review
Letters} \textbf{91}, 066403 (2003).

35. Ulrich, C., G$\ddot{o}$ssling, A., Gr$\ddot{u}$ninger, M., Guennou, M., Roth, H., Cwik, M., Lorenz, T., Khaliullin, G. \& Keimer, B. Raman Scattering in the Mott
Insulators LaTiO$_3$ and YTiO$_3$: Evidence for Orbital Excitations. \emph{Physical Review Letters} \textbf{97}, 157401 (2006).

36. Chien, Chia-Ling \& De Benedetti, S., Barros, F. De S. Magnetic properties of EuTiO$_3$, Eu$_2$TiO$_4$ \& Eu$_3$Ti$_2$O$_7$, \emph{Physical Review B} \textbf{10}, 3913 (1974).

37. Katsufuji, T. \& Tokura, Y. Transport and magnetic properties of a ferromagnetic metal: Eu$_{1-x}$R$_{x}$TiO$_3$, \emph{Physical Review B} \textbf{60}, 15021 (1999).

38. Katsufuji, T. \& Takagi, H. Coupling between magnetism and dielectric properties in quantum paraelectric EuTiO$_{3}$ \emph{Physical Review B} \textbf{64}, 054415 (2001).

39. Takahashi, K. S., Onoda, M., Kawasaki, M., Nagaosa, N. \&  Tokura, Y. Control of the Anomalous Hall Effect by Doping in Eu$_{1-x}$La$_{x}$TiO$_3$ Thin Films, \emph{Physical Review Letters} \textbf{103}, 057204 (2009).

40. Reuvekamp, P., Caslin, K., Guguchia, Z., Keller, H., Kremer, R. K, Simon, A., K\"{o}hler, J. \& Bussmann-Holder, A. Tiny cause with huge impact: polar instability through strong magneto-electric-elastic coupling in bulk EuTiO$_{3}$, \emph{J. Phys.: Condens. Matter} \textbf{27}, 262201 (2015).

41. Kususe, Y., Murakami, H., Fujita, K., Kakeya, I., Suzuki, M., Murai, S. \& Tanaka, K. Magnetic and transport properties of EuTiO$_3$ thin films doped with $Nb$, \emph{Japanese Journal of Applied Physics} \textbf{53}, 05FJ07 (2014).

42. Ishikawa, K., Adachi, Gin-ya \& Shiokawa, J. Electrical Properties of Sintered EuTiO$_3$$-$ EuNbO$_3$. \emph{Mat. Res. Bull.} \textbf{18},
257$-$262 (1983).

43. Roy, S., Khan, N. \& Mandal, P. Giant low-field magnetocaloric effect in single-crystalline EuTi$_{0.85}$Nb$_{0.15}$O$_3$, \emph{APL Materials} \textbf{4}, 026102 (2016).

44. Li, L., Morris, J., Koehler, M., Dun, Z., Zhou, H., Yan, J., Mandrus, D., Keppens, V. Structural and magnetic phase transitions in EuTi$_{1-x}$Nb$_{x}$O$_3$, \emph{Physical Review B} \textbf{92}, 024109 (2015).

45. Li, L., Zhou, H., Yan, J., Mandrus, D. \& Keppens, V. Research Update: Magnetic phase diagram of EuTi$_{1-x}$B$_{x}$O$_3$ ($B$= Zr, Nb), \emph{APL Materials} \textbf{2}, 110701 (2014).

46. Zubkov, V. G., Tyutyunnik, A. P., Pereliaev, V. A., Shveikin, G. P., Kohler, J., Kremer, R. K., Simon, A. \& Svensson, G. Synthesis and
structural,magnetic and electrical characterisation of the reduced oxoniobates BaNb$_8$O$_{14}$, EuNb$_8$O$_{14}$, Eu$_2$Nb$_5$O$_9$ and
Eu$_x$NbO$_3$ (x=0.7,1.0). \emph{Journal of Alloys and Compounds} \textbf{226}, 24 (1995).

47. Kittel C. \emph{Introduction to Solid State Physics} (Wiley) 2005.

48. Madduri, P. V. P.  \& Kaul, S. N. Magnon-induced interband spin-flip scattering contribution to resistivity and magnetoresistance
in a nanocrystalline itinerant-electron ferromagnet: Effect of crystallite size. \emph{Physical Review B} \textbf{95}, 184402 (2017).

49. Mills, D. L., Fert, A. \& Campbell, I. A. Temperature Dependence of the Electrical Resistivity of Dilute Ferromagnetic Alloys, \emph{Physical Review B} \textbf{4}, 196 (1971).

50. Blasco, J. \&  Garcia, J. A comparative study of the crystallographic, magnetic and electrical properties of the Nd$_{1-x}$La$_x$NiO$_{3-\delta}$ system. \emph{J. Phys.: Condens. Matter} 6,  10759 (1994).

51. Wang,  L. M., Horng, H. E. \& Yang, H. C. Anomalous magnetotransport in SrRuO$_3$ films: A crossover from Fermi-liquid to non-Fermi-liquid behavior, \emph{Physical Review B} \textbf{70}, 014433 (2004).

52. Krellner, C., Kini, N. S., Br$\ddot{u}$ning, E. M., Koch, K., Rosner, H., Nicklas, M., Baenitz, M. \& Geibel, C. CeRuPO: A rare example of a
ferromagnetic Kondo lattice, \emph{Physical Review B} \textbf{76}, 104418 (2007).

53. Nikiforov, V. N., Baran, M., Jedrzejczak, A. \& Irkhin, V. Y. Anomalous ferromagnetism and non-Fermi-liquid behavior in the Kondo lattice CeRuSi$_2$ \emph{Eur. Phys. J. B} \textbf{86}, 238 (2013).

54. Hiranaka, Y., Nakamura, A., Hedo, M., Takeuchi, T., Mori, A., Hirose, Y., Mitamura, K., Sugiyama, K., Hagiwara, M., Nakama, T. \& $\bar{O}$nuki, Y. Heavy Fermion State Based on the Kondo Effect in EuNi$_2$P$_2$, \emph{Journal of Physical Society of Japan} \textbf{82}, 083708 (2013).

55. Nakamura, A., Okazaki, T., Nakashima, M., Amako, Y., Matsubayashi, K., Uwatoko, Y., Kayama, S., Kagayama, T., Shimizu, K., Uejo, T., Akamine, H., Hedo, M., Nakama, T. \& $\bar{O}$nuki, Y. \& Shiba, H. Pressure-Induced Valence Transition and Heavy Fermion State in Eu$_2$Ni$_3$Ge$_5$ and
EuRhSi$_3$. \emph{Journal of Physical Society of Japan} \textbf{84}, 053701 (2015).

\textbf{Acknowledgement:}{\scriptsize  We thank R. Singha, S. Roy, M. Das, A. Pariari and A. Paul for their help during measurements and useful discussions.}\\

\textbf{Author Contributions:} {\scriptsize S.R. prepared the sample. S.R. \& N.K performed the experiments. P.M., N.K. and S.R. analysed and interpreted the data. P.M., N.K. and S.R. wrote the paper. P.M. supervised the project.}
\newpage
\pagebreak
\begin{center}
\large \textbf{Supplementary information for "Unconventional transport properties of an itinerant ferromagnet: EuTi$_{1-x}$Nb$_{x}$O$_3$ ($x$=0.10$-$0.20)"}
\end{center}
\begin{center}
\textbf{X-ray Characterization technique}\\
\end{center}
The phase purity of EuTi$_{1-x}$Nb$_{x}$O$_3$ ($x$=0.10$-$0.20) was checked by powder X-ray diffraction (XRD) method with Cu $K_{\alpha}$ radiation at room temperature in a high-resolution Rigaku TTRAX II diffractometer. No trace of impurity phase has been detected within the resolution ($\sim$ $2\% $) of XRD which is shown in Figure 1. It confirms the single phase nature of these samples. The structural analysis of powdered single crystal has been done with the Rietveld refinement method. All the peaks in XRD were well fitted to cubic structure of space group $pm\bar{3}m$ and the observed lattice parameters are quite comparable with the previously reported [1] values for EuTi$_{1-x}$Nb$_{x}$O$_3$.
\begin{center}
\textbf{Magnetization}\\
\end{center}
The temperature dependence of dc susceptibility was measured for all the studied ETNO compounds ($x$=0.10$-$0.20) at magnetic field 50 Oe which is shown in Figure 2. It is clear from the figure that these samples undergo paramagnetic to ferromagnetic transition at Curie temperatures $T_{\rm C}$= 8, 9.5 and 6 K for $x$=0.10, 0.15 and 0.20 respectively.

\begin{center}
\textbf{Heat Capacity}\\
\end{center}
The heat capacity measurement was carried out by conventional thermal relaxation technique using physical property measurement system (Quantum Design) [2]. The temperature dependence of heat capacity is shown for all the studied compounds of ETNO in Figure 3. Each of the three compounds show $\lambda$-like peak just near $T_{\rm C}$ which indicates that the paramagnetic to ferromagnetic phase transition is continuous in nature. No other anomaly coming from the impurity has been observed in heat capacity.
\begin{center}
\textbf{Temperature dependence of resistivity for EuTi$_{1-x}$Nb$_{x}$O$_3$ ($x$=0.10, 0.20)}\\
\end{center}
 The thermal variation of resistivity of the ETNO compounds $x$ = 0.10 and $x$ = 0.20 is shown in Figures 4 and 5 respectively. Similar to EuTi$_{0.85}$Nb$_{0.15}$O$_3$, the resistivity for both of the compounds obeys different power law of temperature for different temperature regimes. They also exhibit $T^2$ and $T^{3/2}$  behaviour in the paramagnetic region above $T_{\rm C}$. But due to having lower magnetic transition temperature we are unable to fit the resistivity of the $x=0.20$ compound with $T^2$ over a reasonable interval of $T$ below $T_{\rm C}$ for the accurate determination of the parameter $a$. Also, as compared to $x$$=$0.10 and 0.15 compositions, the Kondo-like behaviour is very weak for this compound.  Therefore as the doping concentration increases the usual magnetic ordering due to RKKY interaction becomes dominating over the non-magnetic Kondo-singlet formation.\\
 {\large \textbf{Supplementary References}}\\
1. Li, L., Zhou, H., Yan, J., Mandrus, D. \& Keppens, V. Research Update: Magnetic phase diagram of EuTi$_{1-x}$B$_{x}$O$_3$ ($B$= Zr, Nb), \emph{APL Materials} \textbf{2}, 110701 (2014).

2. Quantum Design. \emph{Physical Properties Measurement Systems, Heat Capacity Option user's Manual, 1085-150, Rev. M6} (USA, 2015).
\newpage
 \begin{figure*}[h!]
\includegraphics[width=0.8\textwidth]{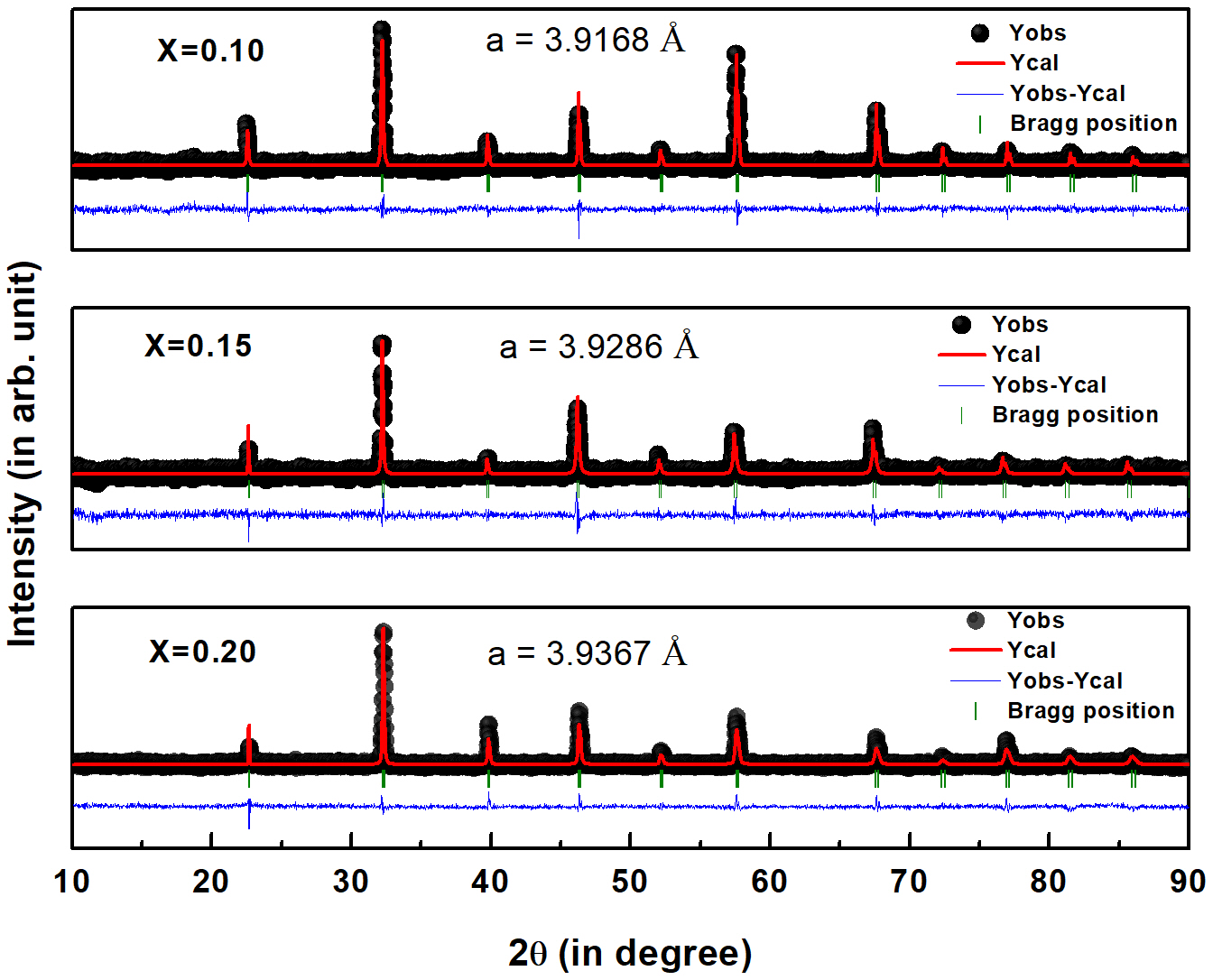}
\renewcommand{\figurename}{\textbf{Supplementary Figure}}
\caption{ \textbf{X-ray powder diffraction pattern for EuTi$_{1-x}$Nb$_{x}$O$_3$ ($x$=0.10$-$0.20) at room temperature. The black solid line corresponds to the Rietveld refinements of the diffraction pattern.} }\label{rh}
\end{figure*}
\begin{figure*}[h!]
\includegraphics[width=0.8\textwidth]{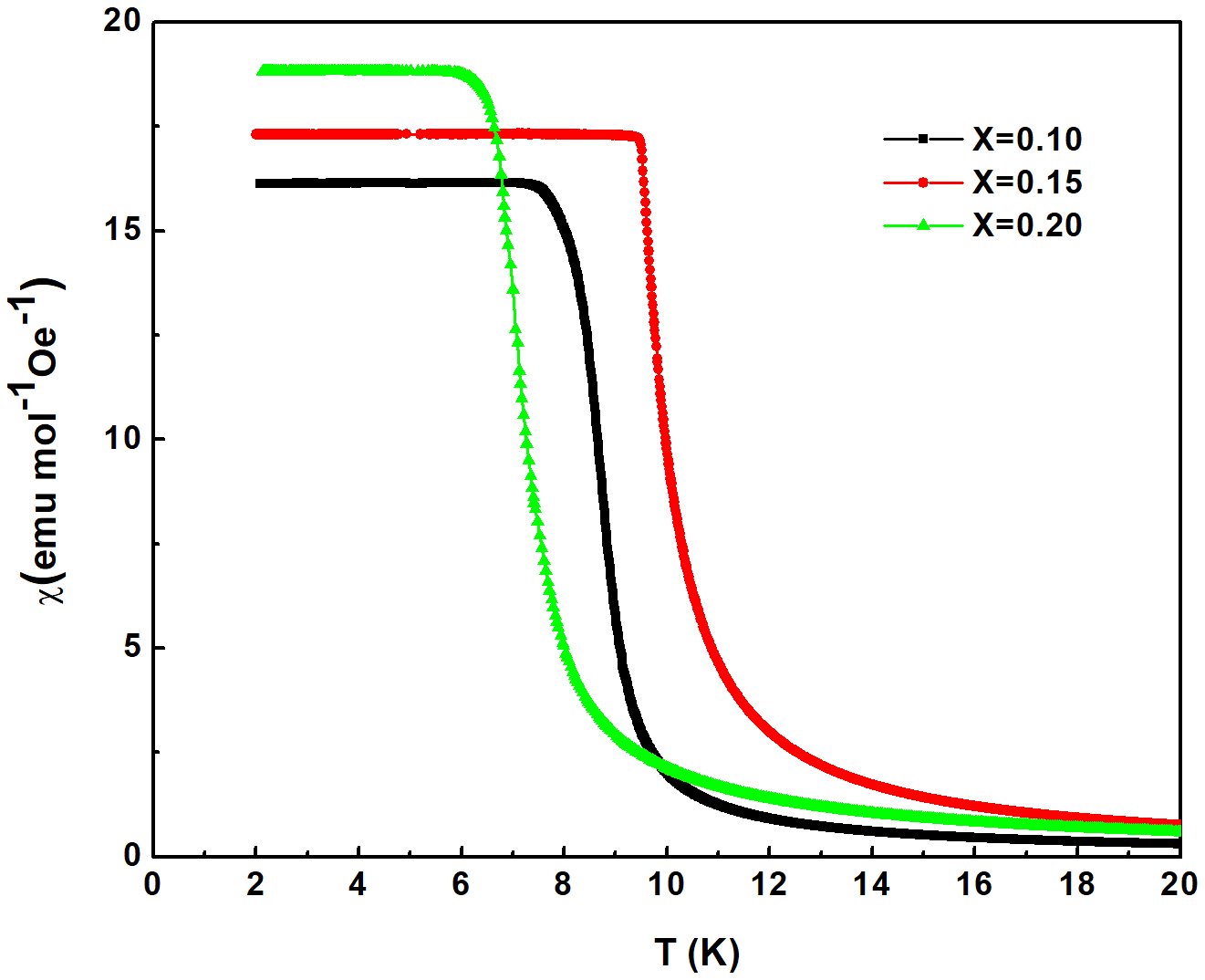}
\renewcommand{\figurename}{\textbf{Supplementary Figure}}
\caption{ \textbf{Thermal variation of dc susceptibility of EuTi$_{1-x}$Nb$_{x}$O$_3$ ($x$=0.10$-$0.20) at 50 Oe magnetic field} }\label{rh}
\end{figure*}
\begin{figure*}[h!]
\includegraphics[width=0.8\textwidth]{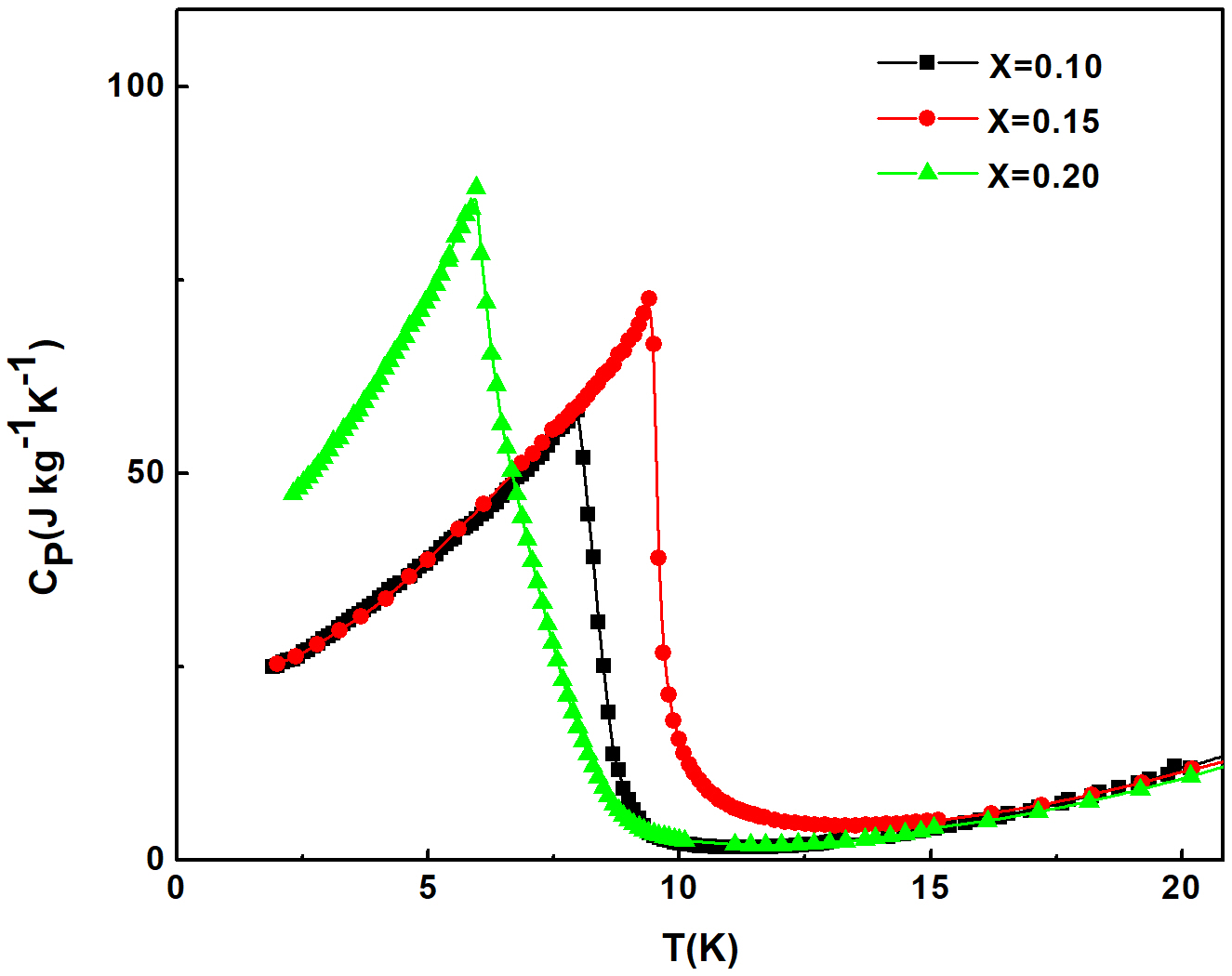}
\renewcommand{\figurename}{\textbf{Supplementary Figure}}
\caption{ \textbf{Thermal variation of heat capacity for EuTi$_{1-x}$Nb$_{x}$O$_3$ ($x$=0.10$-$0.20)} }\label{rh}
\end{figure*}

\begin{figure*}[h!]
\includegraphics[width=0.8\textwidth]{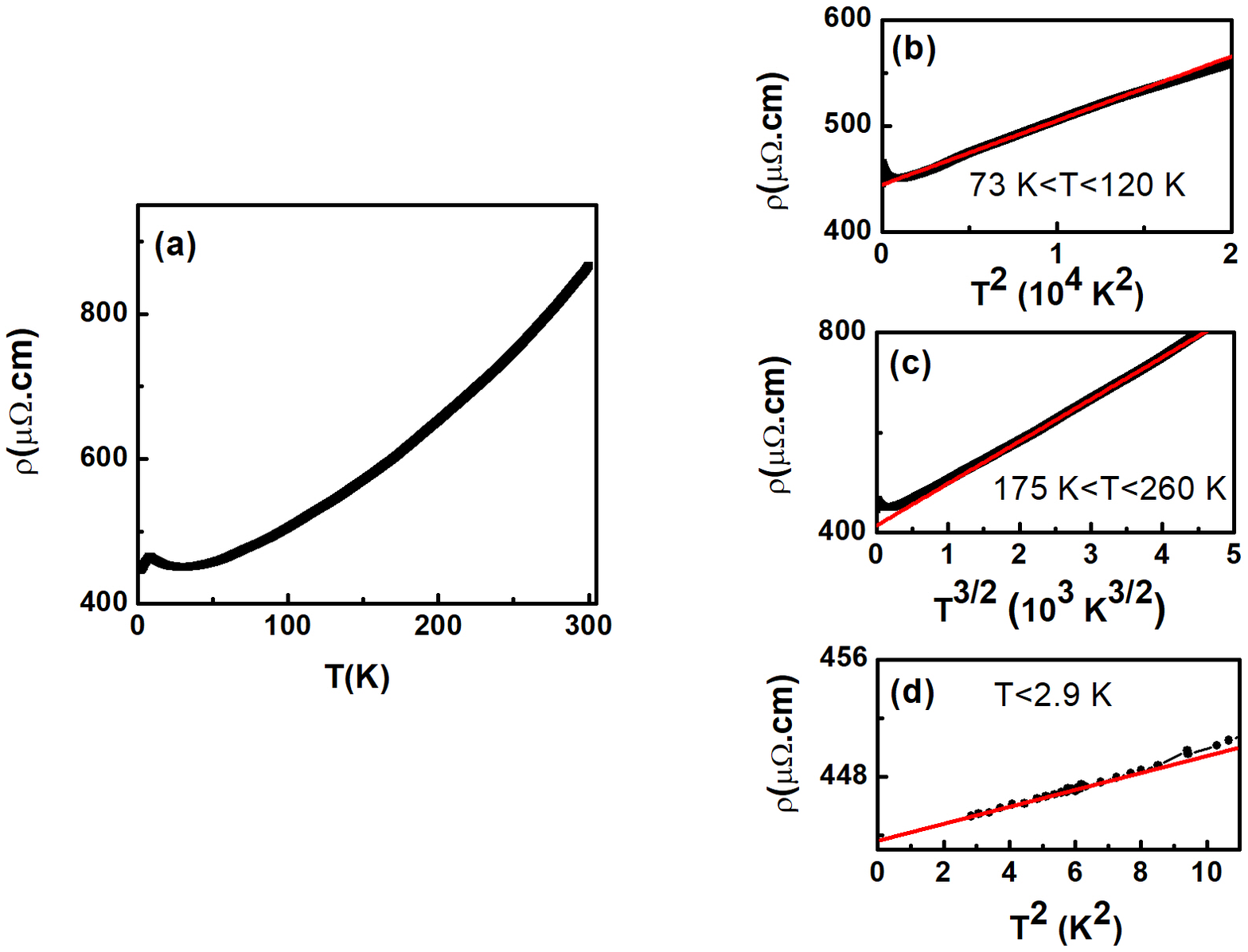}
\renewcommand{\figurename}{\textbf{Supplementary Figure}}
\caption{ \textbf{(a) Temperature dependence of the resistivity of the EuTi$_{1-x}$Nb$_{x}$O$_3$ ($x$=0.10) compound over the whole temperature range. (b) $T^{2}$ dependence of resistivity in the in the paramagnetic region in the temperature range between 73 K to 120 K. (c) $T^{3/2}$ variation of resistivity in the temperature range of 175 K $<$ $T$ $<$ 260 K above T$_{\rm C}$ and (d) $T^{2}$ fitting into the resistivity at very low temperature in the FM region below 2.9 K.}}\label{rh}
\end{figure*}
\begin{figure*}[h!]
\includegraphics[width=0.8\textwidth]{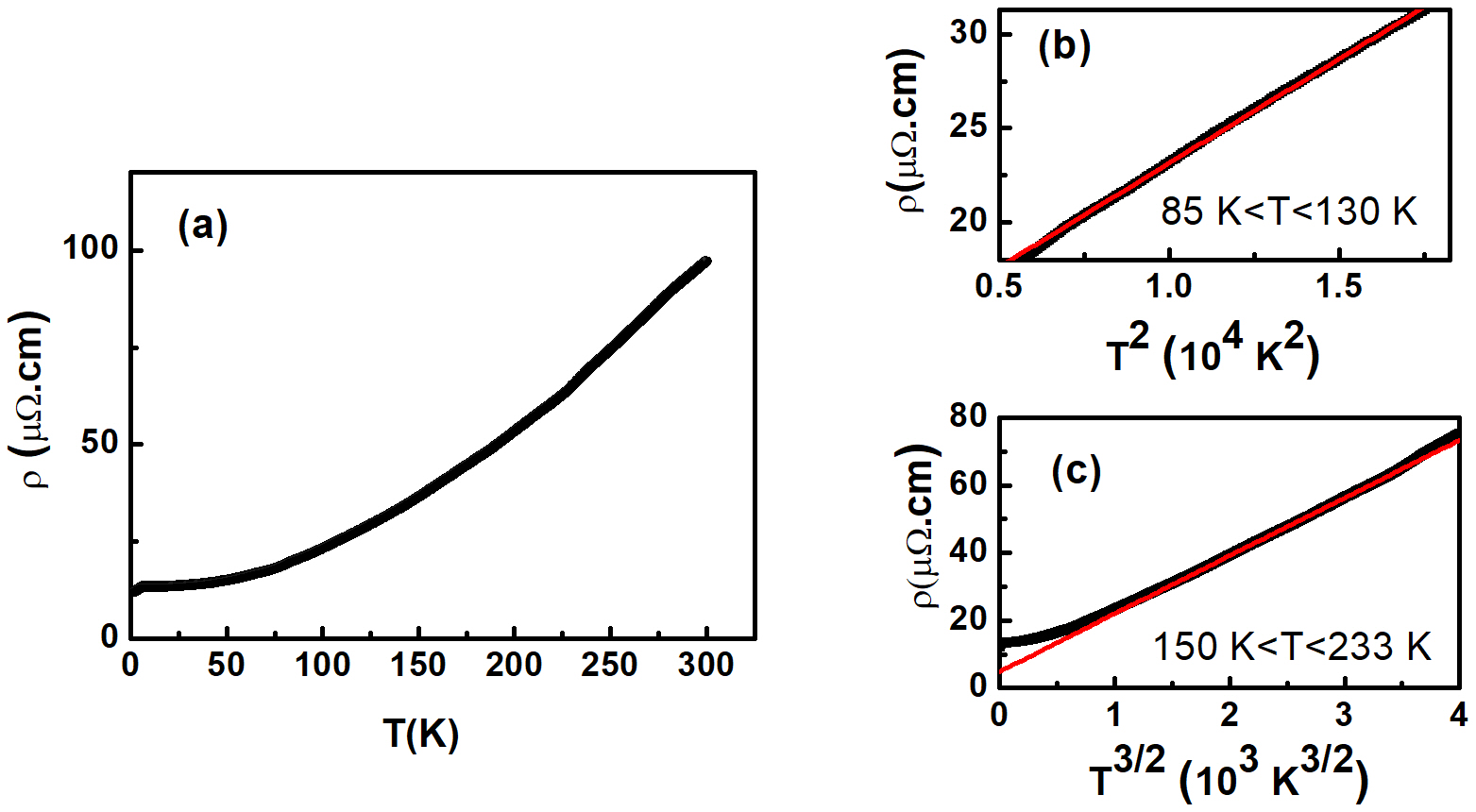}
\renewcommand{\figurename}{\textbf{Supplementary Figure}}
\caption{ \textbf{(a) Temperature dependence of the resistivity of the EuTi$_{1-x}$Nb$_{x}$O$_3$ ($x$=0.20) compound over the whole temperature range. (b) $T^{2}$ dependence of resistivity in the temperature range of 85 K $<$ $T$ $<$ 130 K above T$_{\rm C}$. (c) $T^{3/2}$ variation of resistivity in the temperature range of 150 K $<$ $T$ $<$ 233 K above T$_{\rm C}$}} \label{rh}
\end{figure*}

\end{document}